\documentclass[12pt]{article}
\usepackage{graphicx}

\newcommand\pubnumber{}
\newcommand\pubdate{\today}

\def\institute{Universit\"at W\"urzburg, Institut f\"ur Theoretische Physik und Astrophysik, 97074 W\"urzburg, Germany}
\def\support{\footnote{Speaker}}

\def\Title#1{\begin{center} {\Large #1 } \end{center}}
\def\Author#1{\begin{center}{ \sc #1} \end{center}}
\def\Address#1{\begin{center}{ \it #1} \end{center}}

\newcommand\pubblock{\rightline{\begin{tabular}{l} \pubnumber\\
         \pubdate  \end{tabular}}}
\newenvironment{Abstract}{\begin{quotation}  }{\end{quotation}}
\newenvironment{Presented}{\begin{quotation} \begin{center} 
             PRESENTED AT\end{center}\bigskip 
      \begin{center}\begin{large}}{\end{large}\end{center} \end{quotation}}
\def\Acknowledgements{\bigskip  \bigskip \begin{center} \begin{large}
             \bf ACKNOWLEDGEMENTS \end{large}\end{center}}

\def\beq{\begin{equation}}
\def\eeq#1{\label{#1}\end{equation}}
\def\eeqn{\end{equation}}

\def\beqa{\begin{eqnarray}}
\def\eeqa#1{\label{#1}\end{eqnarray}}
\def\eeqan{\end{eqnarray}}

\let\bar=\overbar

\def\Dslash{\not{\hbox{\kern-4pt $D$}}}
\def\dslash{\not{\hbox{\kern-2pt $\del$}}}

\def\msb{{\bar{\ssstyle M \kern -1pt S}}}

\begin{document}
\begin{titlepage}
\pubblock

\vfill
\Title{NLO QCD corrections to the production of top--antitop pairs\\[.8ex] in
  association with a W boson including leptonic decays}
\vfill
\Author{Ansgar Denner, Giovanni Pelliccioli\support}
\Address{\institute}
\vfill
\begin{Abstract}
We present the QCD radiative corrections to the full off-shell $\rm t\bar{t}W^+$ production,
considering a final state with three charged leptons, two b~jets and missing energy.
All interferences, off-shell effects and spin correlations are included in the calculation. Beyond presenting integrated and
differential results for the full off-shell process, we compare them with those
obtained applying a double-pole approximation to the virtual corrections.
\end{Abstract}
\vfill
\begin{Presented}
$13^\mathrm{th}$ International Workshop on Top Quark Physics\\
Durham, UK (videoconference), 14--18 September, 2020
\end{Presented}
\vfill
\end{titlepage}
\def\thefootnote{\fnsymbol{footnote}}
\setcounter{footnote}{0}

\section{Introduction}
An accurate theoretical modelling is needed for the top-antitop production
in association with a $\rm W$ boson at the LHC. Beyond its own importance both
as a probe of the Standard Model (SM) and as a window to new physics effects,
such a process represents a relevant background to the $\rm t\bar{t}H$ production.

Recent experimental results show sizeable deviations from the SM in the modelling
of $\rm t\bar{t}W$ production \cite{ATLASandCMS}, and much effort is being put to improve
the theoretical description of the process in order to compare with present and future
LHC data.
The next-to-leading-order (NLO) corrections (QCD and electroweak) are known since several years for the inclusive
$\rm t\bar{t}W$ production both for on-shell top quarks and in the narrow-width approximation
\cite{Campbell:2012dh},  and the matching to parton-shower \cite{Garzelli:2012bn}, soft-gluon resummation \cite{Broggio:2019ewu}
and multi-jet merging \cite{vonBuddenbrock:2020ter} have also been tackled for this process.

The first full off-shell calculations of $\rm t\bar{t}W$ production in the three-lepton decay
channel have appeared very recently \cite{Bevilacqua:2020pzy,Denner:2020hgg}.
In these proceedings, we briefly
describe the calculation of Ref.~\cite{Denner:2020hgg} and present the most relevant results
at the integrated and differential level, including a comparison between full off-shell results
and those obtained with a double-pole approximation.

\section{Description of the calculation}\label{sec:descr}
\begin{figure}[b]
  \centering
  \includegraphics[height=1.33in]{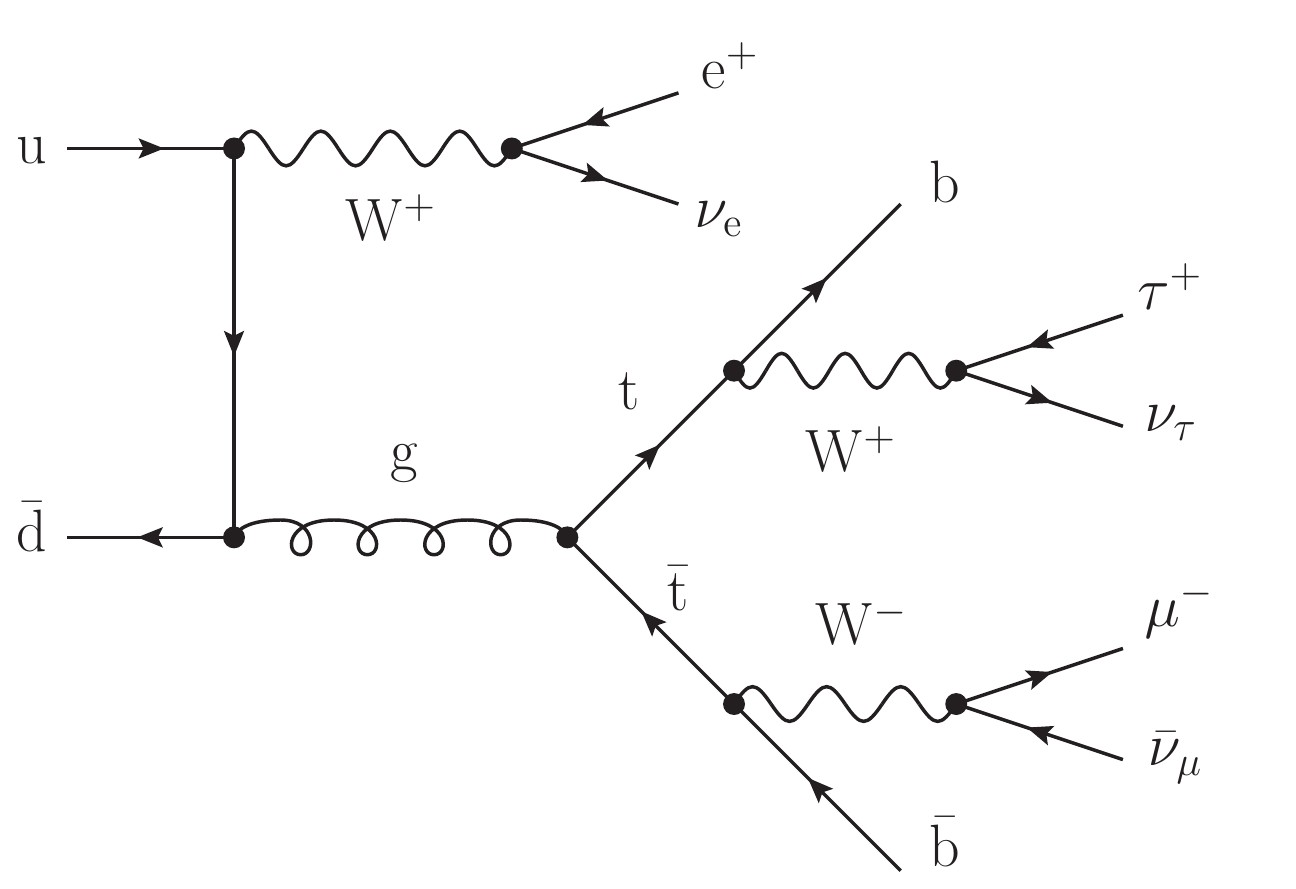}
  \includegraphics[height=1.33in]{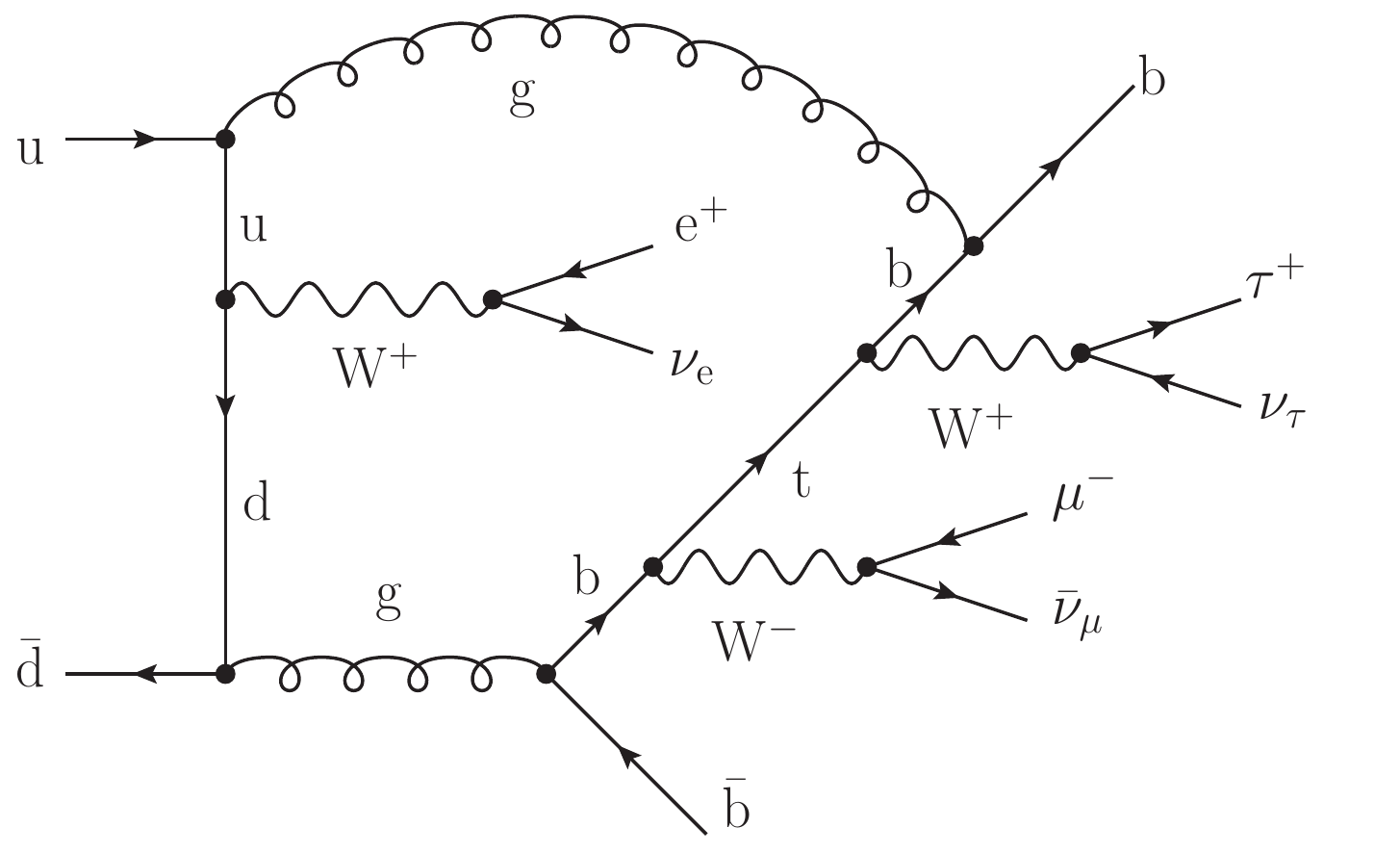}
  \caption{Sample tree-level (left) and one-loop (right) Feynman diagrams.}
  \label{fig:diags}
\end{figure}

We consider the complete NLO QCD corrections to off-shell $\rm t\bar{t}W^+$ production at the LHC@13TeV in the channel
${\rm p}{\rm p}\to {\rm e}^+\nu_{\rm e}\,\mu^-\overline{\nu}_\mu\,\tau^+\nu_\tau\,{\rm b}\,\overline{\rm b}$,
including all off-shell effects, spin-correlations and interferences. At leading-order 
[LO, $\mathcal{O}(\alpha_{\rm s}^2\alpha^6)$], this process receives
contributions from quark-induced partonic channels only.
At NLO QCD [$\mathcal{O}(\alpha_{\rm s}^3\alpha^6)$], also the ${\rm g}q/{\rm g}\bar{q}$ channels open up.
While the virtual corrections feature up to seven-point one-loop functions to be evaluated (see Fig.~\ref{fig:diags}),
the challenging part of the calculation are the real corrections, as they involve a large-multiplicity final state
($2\rightarrow9$ process, with four external coloured partons).
Tree-level and one-loop amplitudes are computed with {\scshape Recola}
\cite{Actis:2016mpe}, and the numerical integration is
performed with the {\scshape MoCaNLO} Monte Carlo code, which uses the
dipole formalism \cite{Catani:1996vz} for
the subtraction of soft and collinear singularities. Unstable particles are treated in the complex-mass scheme \cite{Denner:1999gp}.
For the details on the SM input parameters and the fiducial setup we refer to Sect.~3.1 of Ref.~\cite{Denner:2020hgg}.

\section{Integrated and differential results} \label{sec:res}
We present the fiducial cross-sections for five different central-scale choices, both fixed and dynamical.
Numerical results are shown in Table~\ref{tab:integ}.
\begin{table}
  \centering
\begin{tabular}{c|ccc}
  central scale & {  LO } & {  NLO QCD} &   $K$-factor\\
\hline\\[-0.3cm]
$\mu_0^{\rm (a)} = M_{\rm t} + {M_{\rm W}}/{2}$   &  $0.2042(1)^{+23.8\%}_{-18.0\%}$   &  $0.2452(7)^{+4.5\%}_{-6.8\%}$ & 1.20\\[0.2cm]
$\mu_0^{\rm (b)} = H_{\rm T}/2$ &  $0.1931(1)^{+23.0\%}_{-17.5\%}$   &  $0.2330(9)^{+4.2\%}_{-6.5\%}$ & 1.21\\[0.2cm]
$\mu_0^{\rm (c)} = H_{\rm T}/3$  &  $0.2175(1) ^{+24.2\%}_{-18.2\%}$   &  $0.2462(8)^{+2.8\%}_{-5.8\%}$ & ${1.13}$\\[0.2cm]
$\mu_0^{\rm (d)} = {\left(M_{\rm T,t}M_{\rm T,\bar{t}}\right)}^{1/2}$  &  $0.1920(1)^{+23.0\%}_{-17.5\%}$   &  $0.2394(6)^{+5.4\%}_{-7.2\%}$ & 1.25\\[0.2cm]
$\mu_0^{\rm (e)} = {\left(M_{\rm T,t}M_{\rm T,\bar{t}}\right)}^{1/2}/2$  &  $0.2360(1)^{+24.9\%}_{-18.7\%}$   &  $0.2535(8)^{+3.4\%}_{-5.2\%}$ & 1.07\\[0.1cm]
\end{tabular}
\caption{Fiducial cross-sections and $K$-factors for different central-scale choices for the
  process ${\rm p}{\rm p}\to {\rm e}^+\nu_{\rm e}\,\mu^-\overline{\nu}_\mu\,\tau^+\nu_\tau\,{\rm b}\,\overline{\rm b}$
  at the LHC@13TeV \cite{Denner:2020hgg}. Numerical errors are shown
  in parentheses, and
  scale uncertainties are given in percentages.}\label{tab:integ}
\end{table}
The dynamical scales are based either on the $H_{\rm T}$ observable, defined as
\begin{equation}
  H_{\rm T} = p_{\rm T, miss}+\sum_{i=\ell, \rm b}p_{{\rm T},i}\,,
\end{equation}
or on the geometrical average of the two top-quark transverse masses,
\begin{equation}
  ({M_{\rm T,t } M_{\rm T,\bar{t}}})^{1/2} = (\sqrt{p^2_{\rm T,t}+M^2_{\rm t}}  \sqrt{p^2_{\rm T,\bar{t}}+M^2_{\rm t}} )^{1/2}\,,
\end{equation}
where the ambiguity in the selection of the top-quark is solved choosing the decay products whose
invariant mass is the closest to top-quark pole mass.

The different scale choices lead to NLO QCD corrections between +7\%
and +25\% to the fiducial cross-section
in the fiducial region. Furthermore the theory uncertainties from renormalization- and factorization-scale
variations are sizeably reduced at NLO: while they are at the 20\% level at LO,
they are reduced to 5\% at NLO QCD.

\begin{figure}[tb]
  \centering
  \includegraphics[height=2.3in]{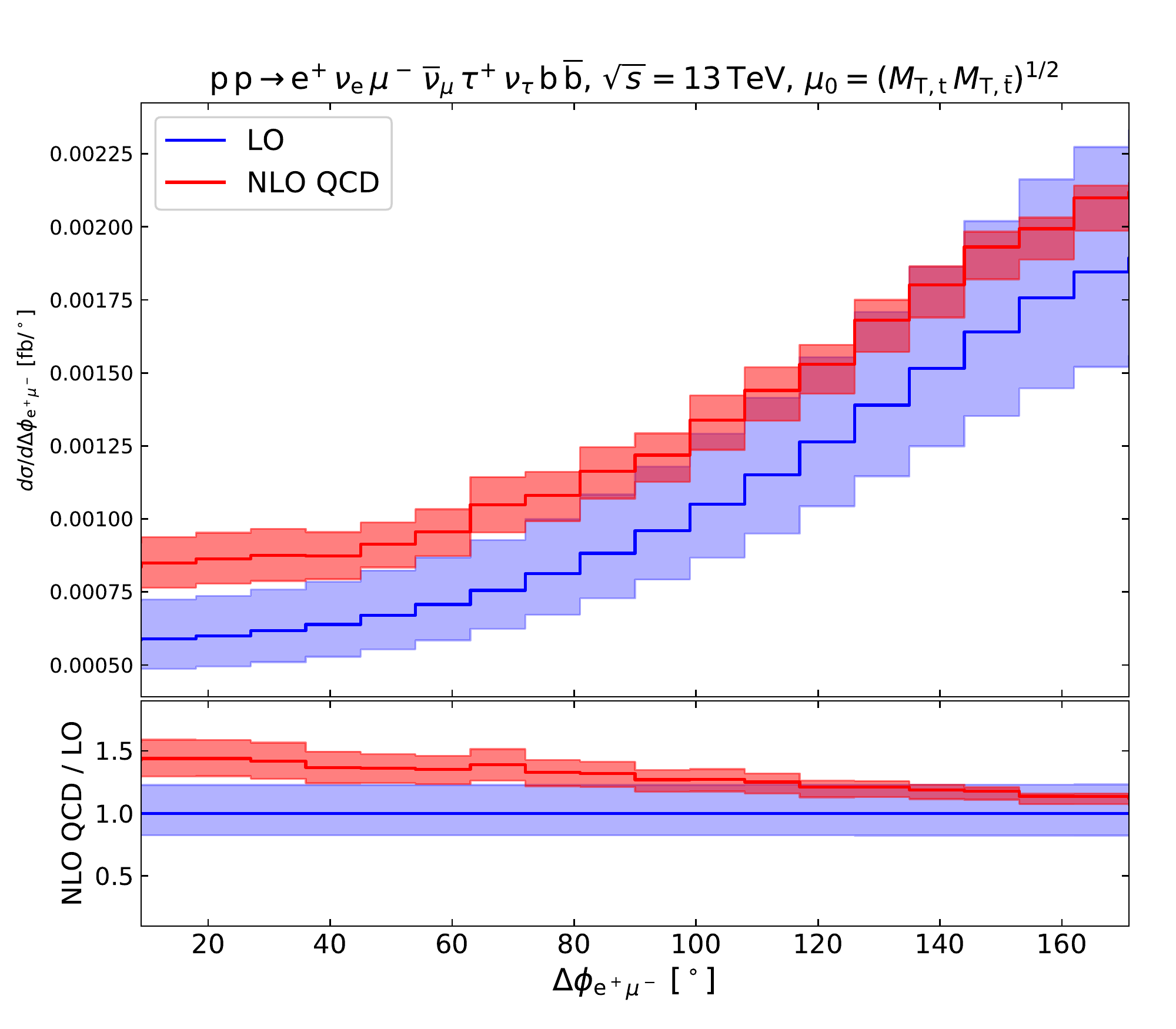}
  \includegraphics[height=2.3in]{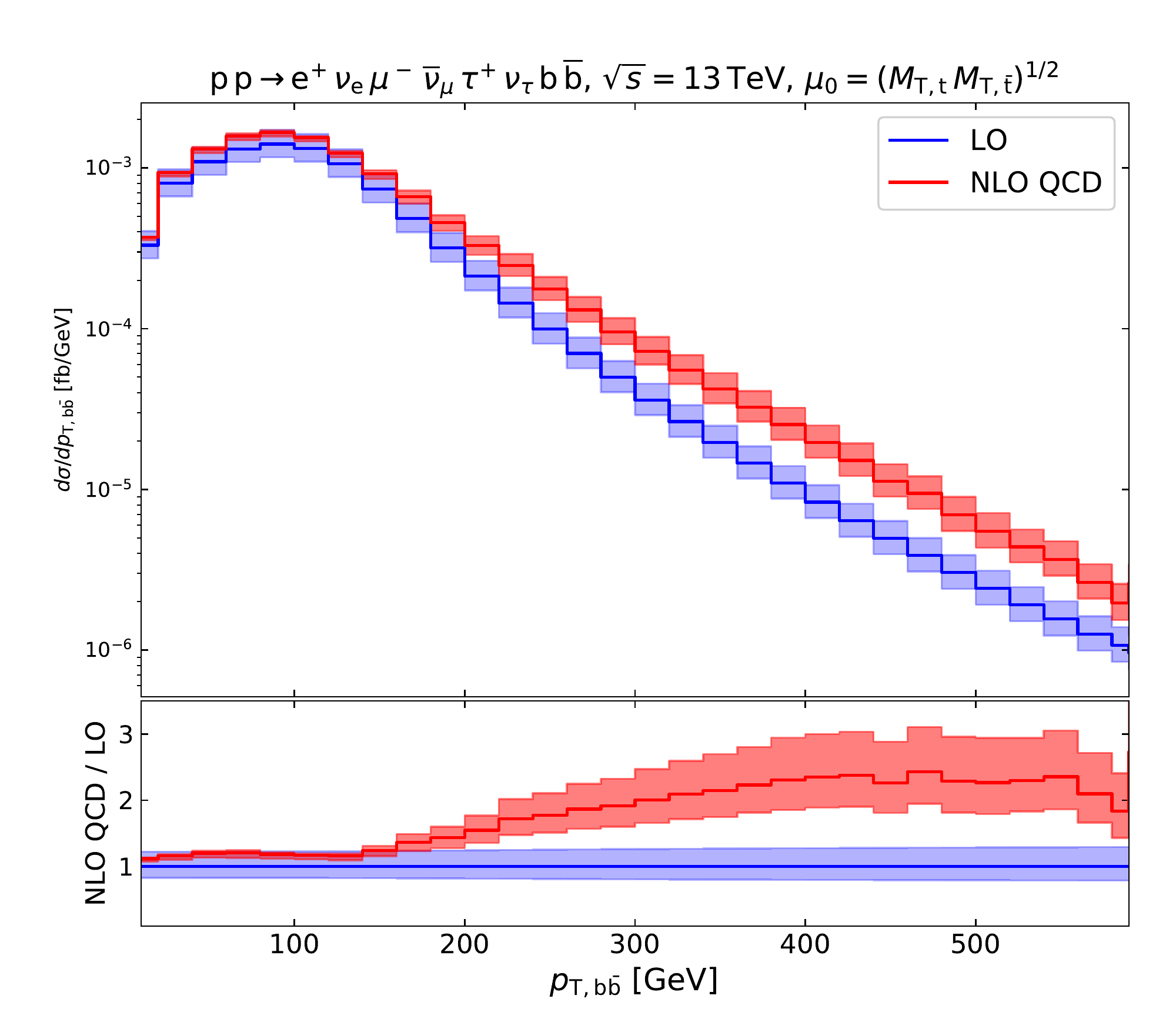}
  \caption{
    Differential cross-sections (upper plot) and $K$-factors (lower plot) for the dynamical scale choice $\mu_0^{\rm (d)}$, for the
    process ${\rm p}{\rm p}\to {\rm e}^+\nu_{\rm e}\,\mu^-\overline{\nu}_\mu\,\tau^+\nu_\tau\,{\rm b}\,\overline{\rm b}$
    at the LHC@13TeV \cite{Denner:2020hgg}.     
    The azimuthal separation between the positron and the muon (left) and the two-b-jet-system transverse momentum (right)
    are considered. Uncertainty bands from 7-point scale variations are shown.}
  \label{fig:diffOFF}
\end{figure}

The choice of a dynamical scale gives flatter corrections than with
a fixed scale at the differential level in most of the analyzed observables.
However, moderate shape distortions owing to NLO corrections can be found
even using a dynamical scale. This is the case for the azimuthal
distance between the positron and the muon, considered in Fig.~\ref{fig:diffOFF} (left).
Despite the usage of the resonance-aware dynamical scale $\mu_0^{\rm (d)}$, the
NLO $K$-factor diminishes from 1.5 (at $\Delta\phi_{\rm e^+\mu^-}\approx 0$) to 1.15 (at $\Delta\phi_{\rm e^+\mu^-}\approx \pi$) .

Another interesting aspect of this process is given by the very large corrections that
characterize some distribution tails, in particular those depending on the b~jets.
As an example, the transverse momentum of the two-b-jet system, shown in Fig.~\ref{fig:diffOFF} (right),
features 100\% corrections already at moderate values.
These effects signal that the NLO QCD corrections are dominated by
real radiation. This is confirmed by the scale-uncertainty bands
that are larger than the LO ones for $p_{\rm T, bb}>200$ GeV.

We have compared the full off-shell results with those obtained
via a double-pole approximation (DPA) for the top-antitop resonances.
More specifically, we have performed two calculations using the DPA:
we have either applied the DPA to the virtual QCD corrections only, or
both to the virtual corrections and to the $I$-operators of the integrated
dipoles, which by construction subtract the explicit infrared poles of the
virtual corrections and therefore should be treated in the same way as the virtual
corrections themselves.

The fiducial cross-sections read (the central scale $\mu_0^{\rm (d)}$ is understood):
\begin{equation}
  \sigma_{\rm NLO}^{\rm full} = 0.2394(6) \rm fb,\qquad \sigma_{\rm NLO}^{\rm DPA,\, V} = 0.2395(7) \rm fb,\qquad \sigma_{\rm NLO}^{\rm DPA,\, V+I} = 0.2422(7) \rm fb\,.
\end{equation}
The agreement between the full off-shell calculation and the one obtained applying the DPA to the virtual
contributions only is impressive, thanks to the fact that virtual
corrections are small for this process. If the DPA is applied also to the $I$-operators of the
integrated dipoles, a 1\% discrepancy is found at the integrated level.

\begin{figure}[tb]
  \centering
  \includegraphics[height=2.3in]{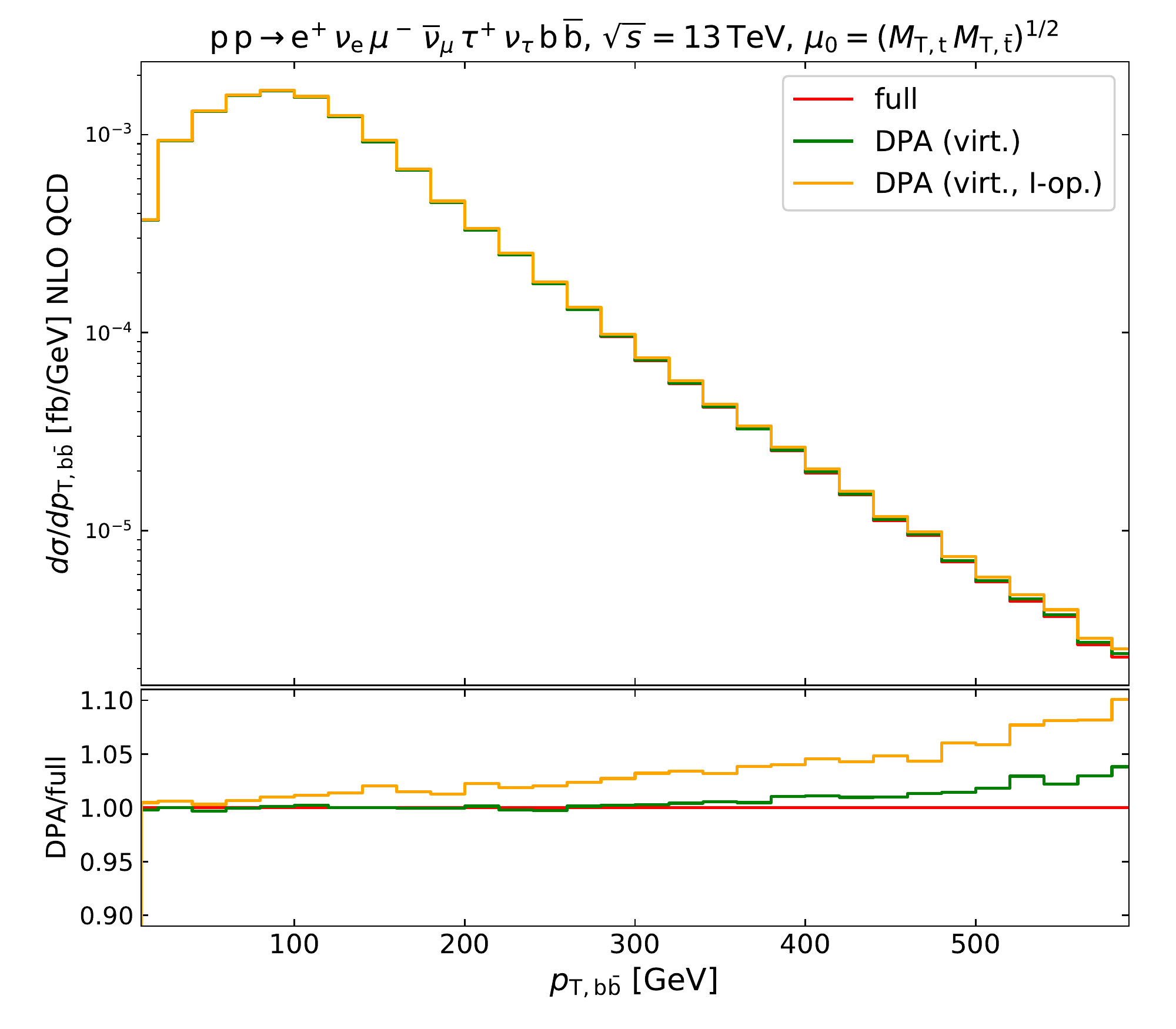}
  \caption{
    Comparison of the off-shell calculation and the one in double-pole approximation for
    ${\rm p}{\rm p}\to {\rm e}^+\nu_{\rm e}\,\mu^-\overline{\nu}_\mu\,\tau^+\nu_\tau\,{\rm b}\,\overline{\rm b}$
    at the LHC@13TeV \cite{Denner:2020hgg}:
    differential NLO QCD cross-sections (top) and ratios over the full
    prediction (bottom) for
  the transverse momentum of the two-b-jet system.}
  \label{fig:diffDPA}
\end{figure}

Larger discrepancies can be observed in more exclusive phase-space regions,
where the contributions of resonant top and antitop quarks are not dominant anymore, and off-shell effects are
sizeable. In Fig.~\ref{fig:diffDPA} we consider again the transverse momentum of the
two-b-jet system. 
If the DPA is applied to the virtual corrections only, the discrepancy w.r.t.\ the
off-shell result reaches 3\% at large values, while 10\% effects appear if also integrated
dipoles are treated in the same way.
The DPA results show that the $\rm t\bar{t}W$ resonant structure is dominant
in most of the fiducial phase space and for sufficiently inclusive observables,
while the full off-shell modelling is definitely needed in suppressed phase-space regions.

\section{Conclusion} \label{sec:concl}
The NLO QCD corrections to  $\rm t\bar{t}W^+$ production at the LHC
are about 20\% for the fiducial cross-section  and sizeably reduce the scale
uncertainties.
Mild differences show up
when using different renormalization and factorization scales.
An improved perturbative convergence is obtained using a dynamical scale,
which gives flatter differential $K$-factors than the typical fixed scale.
Very large $K$-factors are found in exclusive phase-space regions,
in particular where the process is dominated by hard QCD radiation.
A comparison of the full off-shell results with those obtained applying
a double-pole approximation to virtual corrections gives a percent agreement
at the level of the fiducial cross-section. Larger discrepancies (up to 10\%) characterize phase-space
regions which are not dominated by the ${\rm t\bar{t}}$ resonance structure.

\Acknowledgements
This work is supported by the German Federal Ministry for
Education and Research (BMBF) under contract no.~05H18WWCA1.


\begin{thebibliography}{99}


\bibitem{ATLASandCMS}
A.~M.~Sirunyan \textit{et al.} [CMS],
Phys. Rev. Lett. \textbf{120} (2018) 231801
[arXiv:1804.02610 [hep-ex]];
M.~Aaboud \textit{et al.} [ATLAS],
Phys. Lett. B \textbf{784} (2018) 173-191
[arXiv:1806.00425 [hep-ex]];
%
[CMS],
CMS-PAS-HIG-17-004;
 [ATLAS],
ATLAS-CONF-2019-045.
 


  
\bibitem{Campbell:2012dh}
J.~M.~Campbell and R.~K.~Ellis,
JHEP \textbf{07} (2012) 052
[arXiv:1204.5678 [hep-ph]];
F.~Maltoni, M.~L.~Mangano, I.~Tsinikos and M.~Zaro,
Phys. Lett. B \textbf{736} (2014) 252-260
[arXiv:1406.3262 [hep-ph]];
R.~Frederix, D.~Pagani and M.~Zaro,
JHEP \textbf{02} (2018) 031
[arXiv:1711.02116 [hep-ph]].


\bibitem{Garzelli:2012bn}
M.~V.~Garzelli, A.~Kardos, C.~G.~Papadopoulos and Z.~Trocsanyi,
JHEP \textbf{11} (2012) 056
[arXiv:1208.2665 [hep-ph]];
F.~Maltoni, D.~Pagani and I.~Tsinikos,
JHEP \textbf{02} (2016) 113
[arXiv:1507.05640 [hep-ph]];
R.~Frederix and I.~Tsinikos,
Eur. Phys. J. C \textbf{80} (2020) 803
[arXiv:2004.09552 [hep-ph]].

\bibitem{Broggio:2019ewu}
A.~Broggio, A.~Ferroglia, R.~Frederix, D.~Pagani, B.~D.~Pecjak and I.~Tsinikos,
JHEP \textbf{08} (2019) 039
[arXiv:1907.04343 [hep-ph]];
A.~Kulesza, L.~Motyka, D.~Schwartl\"ander, T.~Stebel and V.~Theeuwes,
Eur. Phys. J. C \textbf{80} (2020) 428
[arXiv:2001.03031 [hep-ph]].

\bibitem{vonBuddenbrock:2020ter}
S.~von Buddenbrock, R.~Ruiz and B.~Mellado,
Phys. Lett. B \textbf{811} (2020) 135964
[arXiv:2009.00032 [hep-ph]].

\bibitem{Bevilacqua:2020pzy}
G.~Bevilacqua, H.~Y.~Bi, H.~B.~Hartanto, M.~Kraus and M.~Worek,
JHEP \textbf{08} (2020) 043
[arXiv:2005.09427 [hep-ph]];
G.~Bevilacqua, H.~Y.~Bi, H.~B.~Hartanto, M.~Kraus, J.~Nasufi and M.~Worek,
[arXiv:2012.01363 [hep-ph]].

\bibitem{Denner:2020hgg}
A.~Denner and G.~Pelliccioli,
JHEP \textbf{11} (2020) 069
[arXiv:2007.12089 [hep-ph]].

\bibitem{Actis:2016mpe}
S.~Actis, A.~Denner, L.~Hofer, J.~N.~Lang, A.~Scharf and S.~Uccirati,
Comput. Phys. Commun. \textbf{214} (2017) 140-173
[arXiv:1605.01090 [hep-ph]].

\bibitem{Catani:1996vz}
S.~Catani and M.~H.~Seymour,
Nucl. Phys. B \textbf{485} (1997) 291-419
[erratum: Nucl. Phys. B \textbf{510} (1998) 503-504]
[arXiv:hep-ph/9605323 [hep-ph]].

\bibitem{Denner:1999gp}
A.~Denner, S.~Dittmaier, M.~Roth and D.~Wackeroth,
Nucl. Phys. B \textbf{560} (1999) 33-65
doi:10.1016/S0550-3213(99)00437-X
[arXiv:hep-ph/9904472 [hep-ph]].

\end{thebibliography}
\end{document}